\documentstyle[12pt,psfig,rotate]{article}
\let\section=\subsection     \let\subsection=\subsubsection                %%

\begin{document}

\begin{center}
   {\large \bf HADRONS IN MEDIUM -- OBSERVABLES\footnote{Work supported by GSI
    Darmstadt and BMBF}}\\[5mm]
   {\large \bf U. Mosel}\\[5mm]
   {\small \it Institut fuer Theoretische Physik, Universitaet Giessen \\
   Heinrich-Buff-Ring 16, D-35392 Giessen, Germany\footnote{email: mosel@theo.physik.uni-giessen.de} \\[8mm] }
\end{center}

\begin{abstract}\noindent
   The observable consequences of a lowering of the restmasses of vector
   mesons on dilepton spectra in the SIS energy regime (HADES) and on
   photoabsorption cross sections on nuclei are studied in a
   transporttheoretical framework. We also investigate the consequences
   of a mass-shift of kaons in dense matter and compare with recent
   SIS data. The need for a detailed
   understanding of the elementary processes is underlined.
\end{abstract}

\section{Introduction}
Predictions of marked changes of hadronic properties in medium have
recently found widespread interest because these predictions are
usually based on some approximation to QCD. Directly QCD-related
effects, such as the predicted partial restoration of chiral symmetry at
high nuclear densities, might thus be observable in dense nuclear systems
\cite{Hatsuda,Weise}. The
overall picture, though, is somewhat confusing since more `classical`
mechanisms, such as meson-baryon couplings also lead to changes of
mesonic properties in the medium \cite{Rapp,Friman}. In addition, all these
calculations
are based on highly idealized scenarios in which infinite nuclear matter
at high density rests forever in an equilibrium state. This assumption
is, of course, never realized in a nuclear reaction, the less so the
more violent the collision is.

We have, therefore, now for several years developed semiclassical
transport theories that allow one to follow the dynamic evolution of a
nuclear reacton, be it between two heavy ions or between a hadron or
photon and a nucleus. In this way we can keep track of the dynamical
evolution of density and temperature during the collision and can
time-locally use the appropriate hadronic properties for the prediction
of observables.

\section{Dileptons}

The results reported here are based on a newly developed transport code
that explicitly propagates all baryonic resonances up to a mass of 2 GeV
and in addition the $\pi, \eta, \rho$  and the $\sigma$, the latter
being a correlated $2 \pi$ pair in an $s$-state \cite{Teis}. It contains
momentum-dependent self-energies for the baryons, none for the mesons
and all resonance decay properties are taken from the particle data
booklet.

The CERES data \cite{Drees} clearly show a large surplus in the invariant
mass
spectrum when compared to socalled `cocktail plots' that are generated
from known decays of produced mesons in nucleon-nucleon collisions
\cite{Drees}. Note, that a very similar effect was already observed by
the DLS group working at the BEVALAC in Berkeley \cite{DLS-Malo}; there
the $\eta$- and the $\pi$-Dalitz decays could not explain the dilepton
spectra in the high-mass region. This deficiency was
explained by us as being due to secondary reactions in which the
produced pions collide to form a $\Delta$ which then decays again
\cite{Batko,Ko,Wolf}. If this same effect is taken into account in
analyzing the CERES data the dramatic discrepancy of a factor 5- 7 is
reduced to about a factor 3, but -- contrary to the DLS case -- does not
dis\-ap\-pear. Dynamical simulations have shown that this surplus can indeed
be explained if the mass of the $\rho$-mesons is decreased according to
the QCD sum rule prediction \cite{CassingKo}.

We have recently shown that at bombarding energies of about 1 - 2 A GeV
the dilepton spectrum contains all the same components as in the CERES
experiment at a much higher energy regime \cite{Bratkovskaya}. The relative
weights are, of
course, somewhat different and the nucleon-nucleon bremsstrahlung,
which proceeds mainly through an excitation of an intermediate $\Delta$
\cite{deJong}, plays an important role at the lower bombarding energy for
invariant masses below about 400 MeV. It is thus essential to know this
component reliably and we have, therefore, recently done microscopic
calculations of this process employing a realistic $T$-matrix that
describes the NN scattering phases up to about 1 GeV \cite{deJong}.

\begin{figure}
\centerline{\psfig{figure=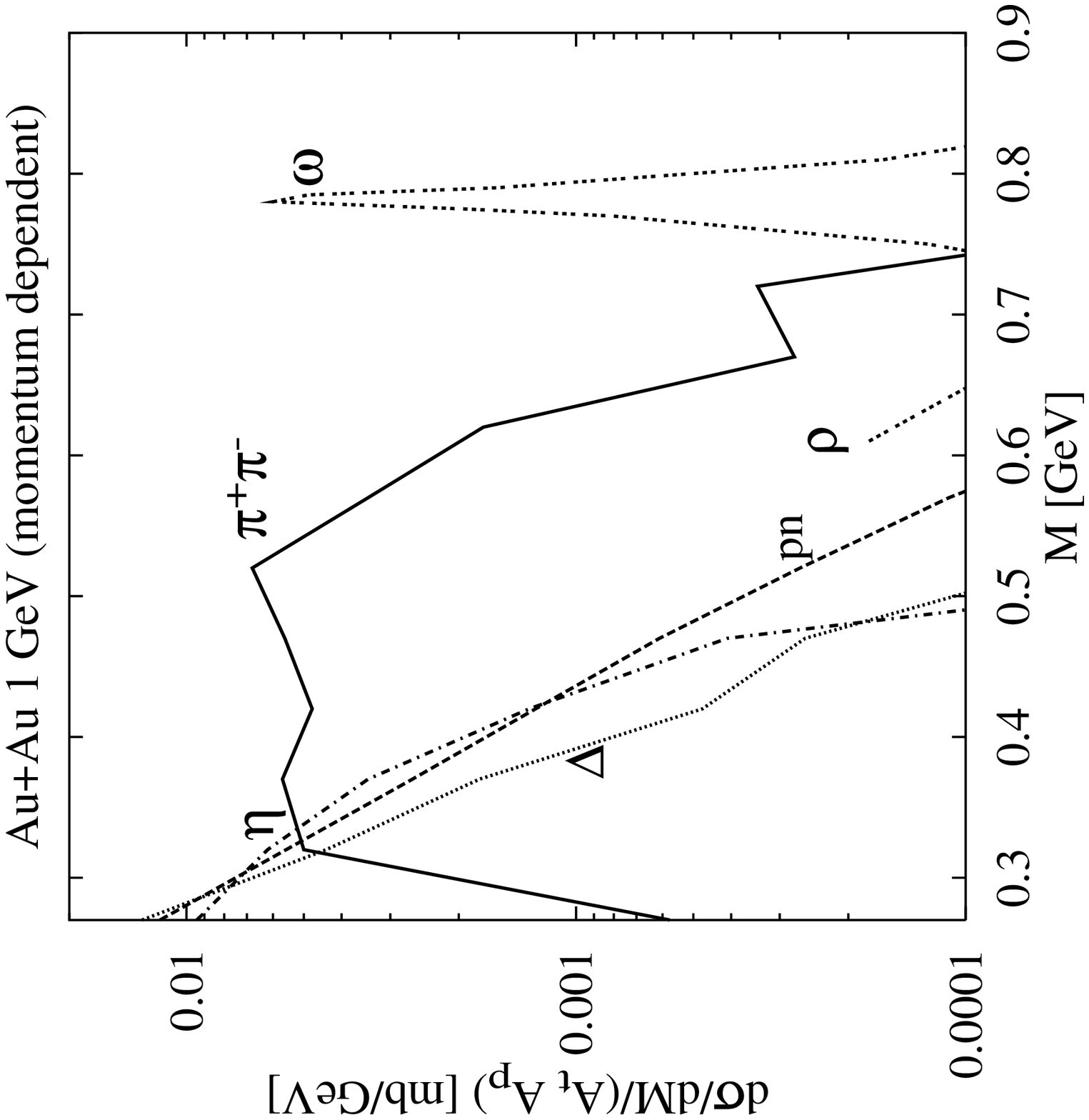,width=80mm}}
\centerline{\psfig{figure=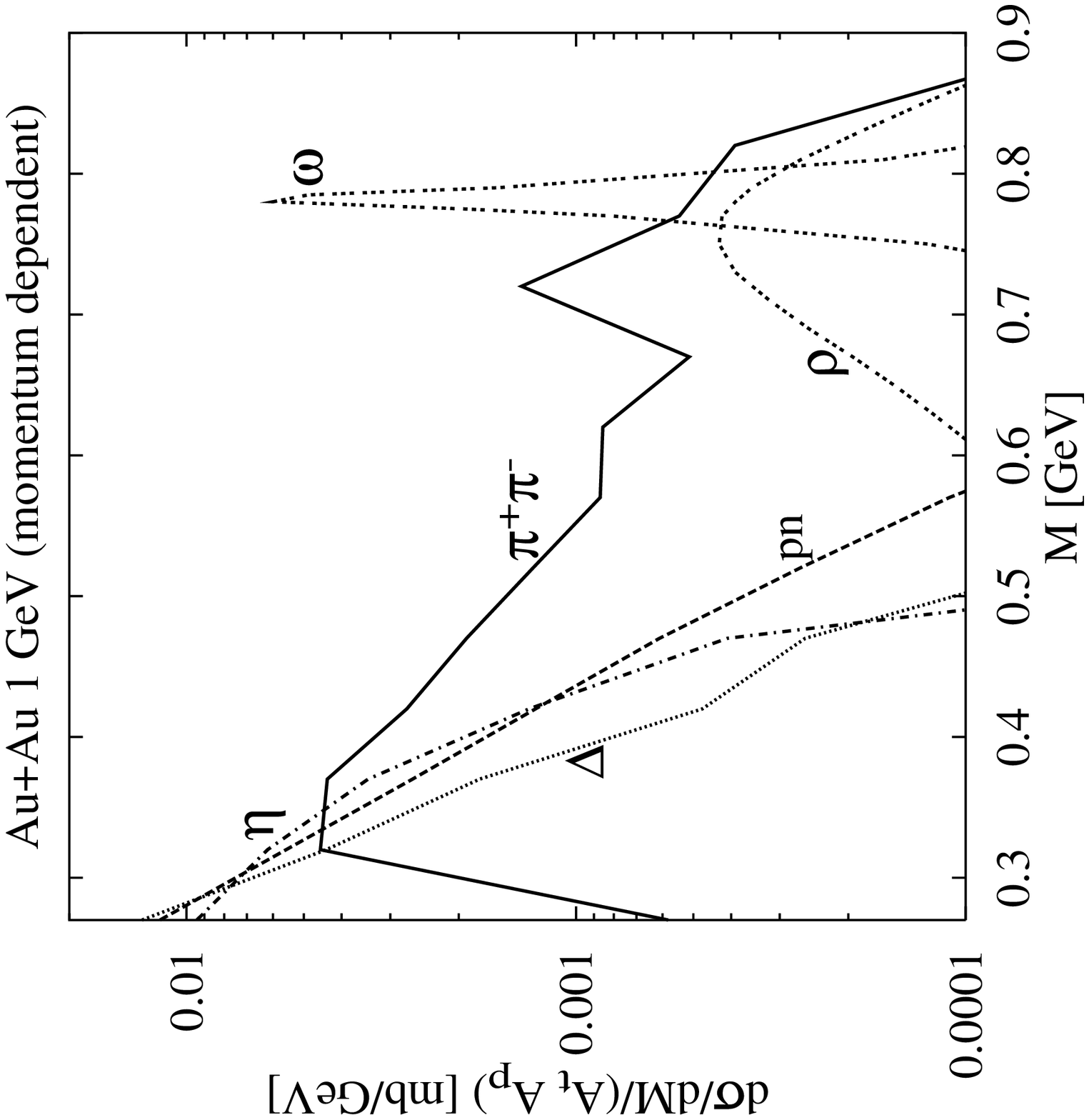,width=80mm}}
\label{Dilept1}
\caption{Invariant mass spectrum for dileptons for $Au + Au$ at 1 AGeV.
Shown are the major components, assuming free (left) and medium-dependent
(right) masses for the vector mesons. The calculations were performed
with the method described in \protect\cite{Teis}, the $\omega$ was not
medium-changed.}
\end{figure}

The full dilepton spectrum calculated with the new transport code,
both without and with a medium-dependent shift of the $\rho$-meson mass
is shown in fig. \ref{Dilept1}.
The in-medium shift of the $\rho$ peak leads to a distinct change of the
spectrum by a factor of 3, which should be measurable by HADES, the new
dilepton spectrometer presently being built for SIS.

\section{Photoabsorption}

An in-medium effect in a completely different type of reaction has
recently generated a lot of interest in the photonuclear community. In
photoabsorption measurements on nuclei it was found that the
higher-lying nucleon-resonances (above the $\Delta$), which are clearly seen
in photoabsorption on the nucleon, dis\-ap\-pear when the same experiment is
performed on a nucleus. This is quite surprising since
the photoabsorption cross section on nuclei scales nearly linearly
with massnumber $A$.

When explaining these data in terms of Breit-Wigner parametrizations for
the individual resonance strengths one is lead to much larger resonance
widths than those found for the nucleon \cite{Alberico,Bianchi}. In addition, also
the overall strength changes somewhat, evidence for shadowing, on the other
hand, sets in only at the highest photon energies \cite{Bianchi}.

\begin{figure}
\centerline{\psfig{figure=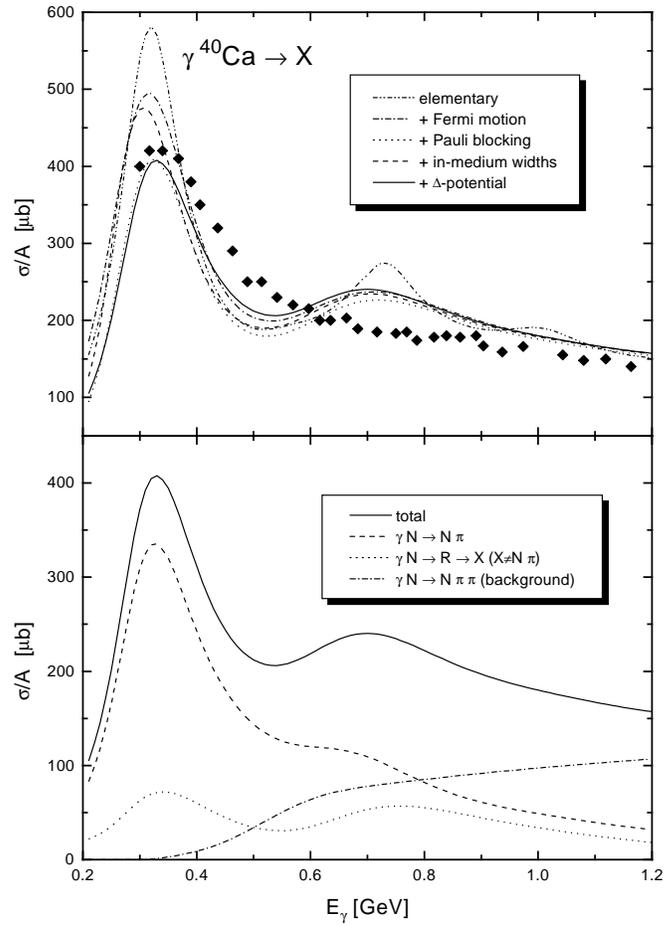,width=9cm}}
\caption{Total photoabsorption cross section. The dots in the upper part give
the data while the lines represent the results of the calculations (for
details see \protect\cite{Effenbg}). The lower part gives a breakdown of the
theoretical cross section into various components.}
\label{photoabsorption}
\end{figure}

\begin{figure}
\centerline{\rotate[r]{\psfig{figure=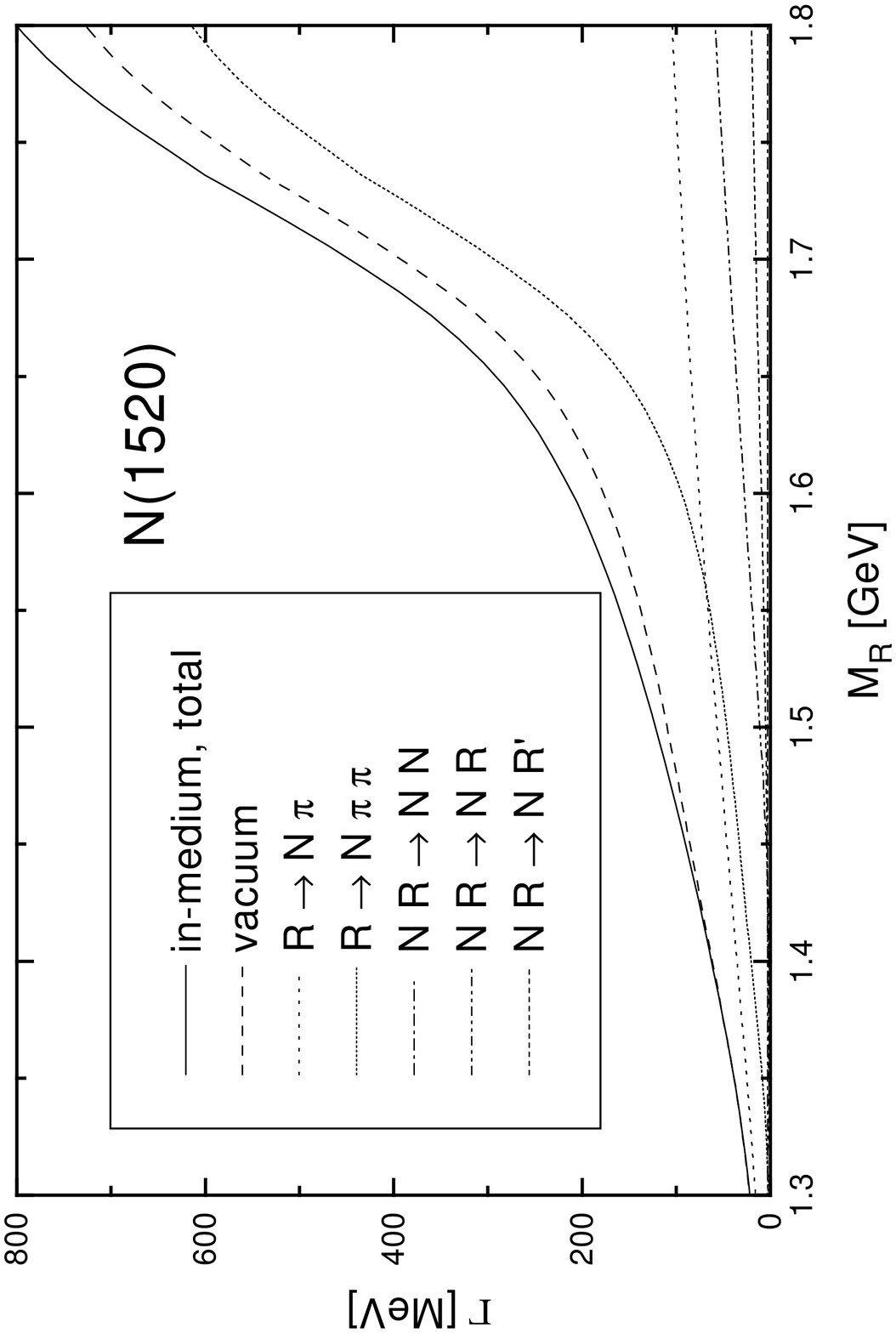,width=8cm}}}
\caption{Energy-dependence of resonance widths. Shown are -- among
others -- the free width, the collision-broadened width and the total
width (from \protect\cite{Effenbg}).}
\label{collbroad}                                     
\end{figure}

We have recently calculated the collisional broadening of the resonances
by using the same transport theory as developed for the description of
heavy-ion resonances \cite{Effenbg} (see fig. \ref{photoabsorption}).
In these studies it turned out
that overall the collisional broadening is not very effective and
clearly not large enough to explain the complete disapppearance of the
resonances (see fig. \ref{collbroad}). It also turned out from these studies that the absence of
resonances beyond masses of about 1600 MeV could be easily understood in
terms of simple Doppler-broadening due to the Fermi-motion which is the
more effective the higher the mass of the resonance is. The real
challenge then is to understand the absence of the N(1520) $D_{13}$
resonance which in the calculation got somewhat broader, but did not
disappear \cite{Effenbg}.

Closer analysis of these results showed, however, that the calculated
`resonance` in the photoabsorption channel coincides with the opening of
the $2\pi$ background production channel. The latter is ununderstood,
even on the nucleon: here the $2\pi$ decay branches of the resonances as
given in the particle data booklet can account only for a small part of
the observed $(\gamma, 2\pi)$ cross section \cite{Effenbg}. The remainder
is then
simply ascribed to a background of unknown origin. If this background
undergoes an in-medium change such that the $2\pi$ background opens up
more slowly, then the `resonance` strength in this region would get
diluted.

A further, speculative, effect is based on the same in-medium mass
change for the $\rho$ meson as used earlier in calculating the dilepton
spectra. The N(1520) resonance has a 20 \% $(N\, \rho)$ decay branch.
Because of the widths of both the resonance and the meson this is
possible, even though the resonance lies about 130 MeV below the
energetic threshold as calculated from the pole masses. If now the
$\rho$ meson is lowered in the nuclear medium by about 18 \%, as
predicted by the sum rule predictions \cite{Hatsuda}, then the
aforementioned decay branch would be energetically open and this would
increase the total resonance width dramatically, as can be seen from
fig. \ref{collbroad}.

\section{Kaon Selfenergies}

Based on an early suggestion of Nelson and Kaplan \cite{Nelson} the in-medium
self\-energy of kaons has recently been reinvestigated \cite{WeiseK}. In
these studies it is found that the restmass of positively charged
kaons ($K^+$) increases slightly with density, while that of negatively
charged antikaons ($K^-$) is predicted to be lowered signigifantly.
An experimental indication for such an effect seemed to be present in
data of the KAOS collaboration at SIS \cite{Kaos} taken at 2
different bombarding energies chosen such that they corresponded to the same
energy above the elementary production threshold. These data show that the $K^-$
production cross sections in heavy-ion collisions are as large as those
for $K^+$ production. In the absence of any in-medium effects this was
unexpected since $K^-$ mesons have a much smaller mean-free-path in
nuclear matter than $K^+$ mesons so that the production of the former
should be surpressed.

First calculations by Ko et al \cite{KoK} have shown that a $K^-$ selfenergy
indeed
has an influence on the production rates. The relative importance of the
various microscopic production channels contributing to the observed
yield depends very much on the microscopic cross-sections used for the
elementary baryon-baryon channel. All the earlier calculations of heavy-ion induced
$K$ production used extrapolations of these cross-sections from
the few measured points several 100 MeV above threshold down to threshold,
since in the most important
region around about 200 MeV above threshold no elementary production data
are available.

Recently, Sibirtsev et al \cite{Sibirtsev} have proposed a new model for
the microscopic
$K^+$ and $K^-$ cross sections that is based on an effective boson
exchange model and has the advantage that the cross sections in the
various isospin channels are constrained by $SU(2)$ symmetry. In
addition, the model automatically includes the correct 4-body to 3-body
phase-space factor for $K^-$ vs. $K^+$ production; this was not the case
in the earlier parametrizations. Very close to threshold (2 MeV above)
the Sibirtsev prediction has found some support recently from the new
COSY-11 measurement; it also describes the older data at higher energy
\cite{BratkovskayaK}.
\begin{figure}
\centerline{\rotate[l]{\psfig{figure=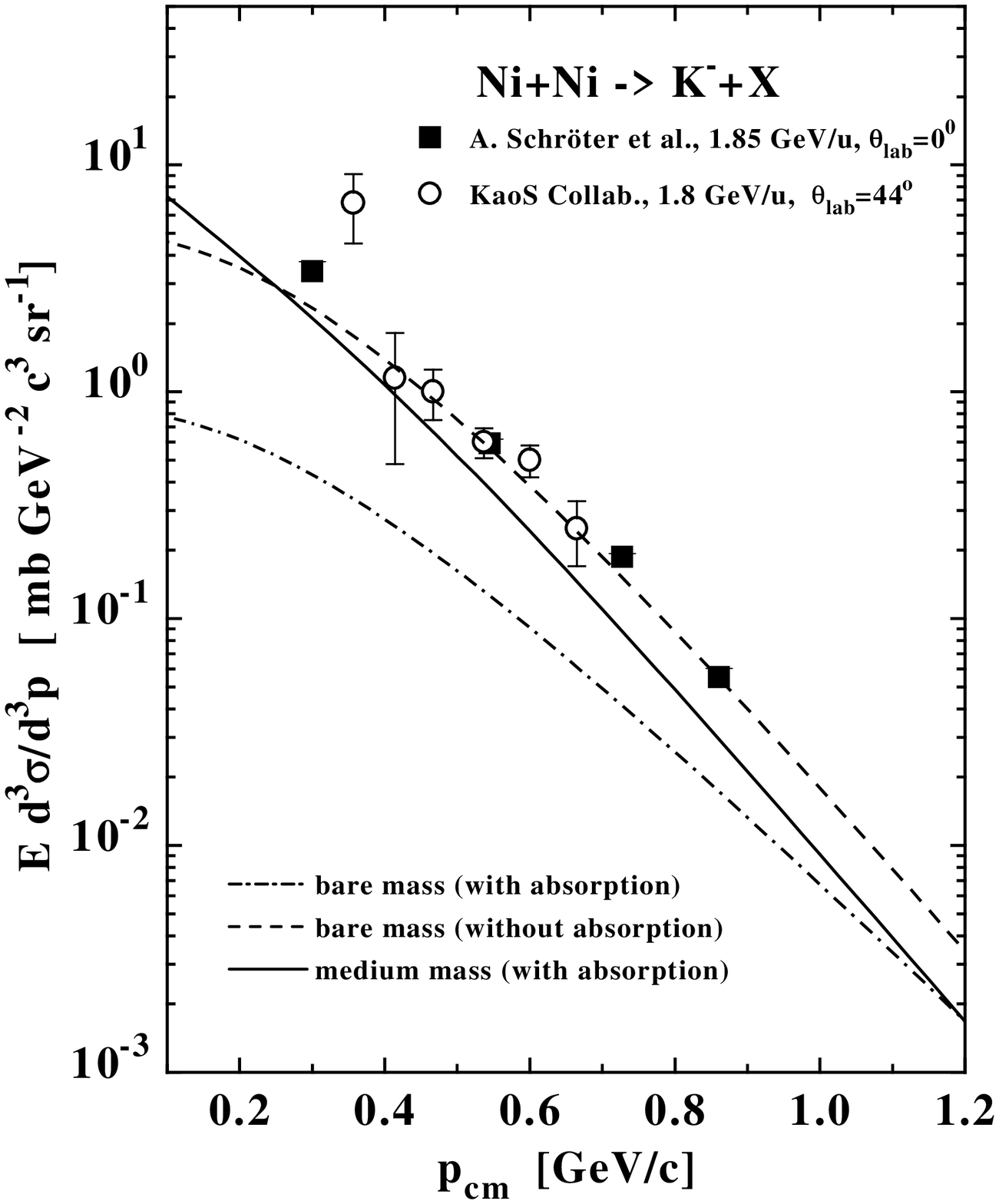,width=80mm}}}
\centerline{\rotate[l]{\psfig{figure=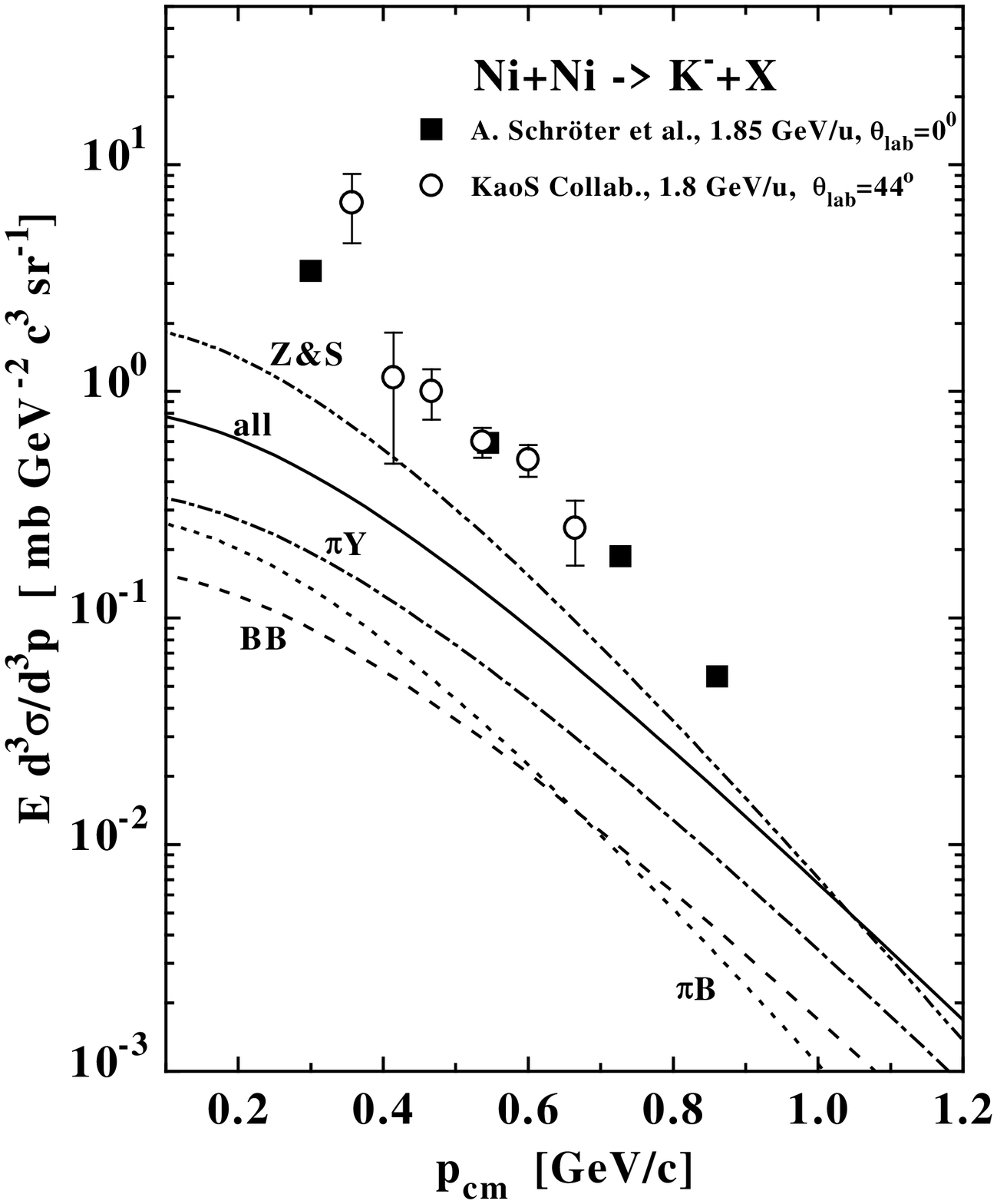,width=80mm}}}
\caption{$K^-$ production cross sections for $Ni + Ni$ at the
         energies indicated. On the left, the solid line gives the sum of all
         production channels, the line labeled $\pi Y$ the contriution
         from pion-hyperon collisions and the curve labeled $Z\&S$ the
         total cross section when the old parametrization for the
         baryon-baryon channel is used. On the right, the solid curve
         gives the final
         result containing both reabsorption and mass-reduction of the
         $K^-$   (from \protect\cite{BratkovskayaK}). }
\label{figK-}
\end{figure}
Using these microscopic cross-sections we have recently reanalyzed the
kaon production by performing calculations in a transport model, that
includes all the relevant kaon-channels \cite{BratkovskayaK}. The result
of these studies is that the $K^+$ channel is quite well described
without using any selfenergy, though a small positive shift cannot be
excluded. The $K^-$ productions rates calculated
without in-medium mass change, on the other hand, lie all below the
measured data (see fig. \ref{figK-}). It is interesting to see that the
dominant production channel is now the $\pi \Sigma$ entrance channel,
i.e. a channel that involves 2 secondary particles, and not the
$\Delta-N$ channel as in earlier studies. This is due to the fact that
the new cross-sections in the baryon-baryon channel are in the relevant
energy range of about 100 - 400 MeV above the elementary threshold
lower than the previously used empirical parametrizations.
However, when including a lowering
of the $K^-$ mass by an amount as predicted by the calculations cited
\cite{Weise} then the cross sections can be reproduced very
satisfactorily (see fig. \ref{figK-}).

\section{Summary and Conclusions}

In this talk I have shown that dilepton spectra in the HADES energy range
between about 1 and 2 GeV are composed of the same elementary sources as
in the relavativistic CERES regime. This observation raises the
expectation that both experiments can complement each other; while the
latter (CERES) reaches higher densities it also leads to higher
temperatures, the former experiment (HADES) works in a `colder`
environment. The effects of predicted in-medium mass-shifts on the
dilepton spectra in the HADES energy range are expected to be as large
as in the CERES regime.

I have also pointed out an interesting possible connection of this
physics problem with the observed disappearance of nucleon resonances in
the photoabsorption cross sections. Here the N(1520) resonance plays an
important role: if the $\rho$ meson mass really drops in nuclear matter
then already at normal nuclear density the decay width of the resonance
into the $N \rho$ channel should increase dramatically, thus helping to
smear out the resonance. A quantitative understanding of the
disappearance of the resonance structures in photoabsorption on nuclei
requires, however, also the investigation of the $2\pi$ channel.

Finally, I have shown new results for the production of positively and
negatively charged kaons. The data by the KAOS collaboration could be
described with no or a slightly positive mass correction for the $K^+$
and a significant lowering of the $K^-$ mass. Since this result depends
crucially on the magnitude of the $K$ production cross sections in the region
of 100 - 400 MeV above threshold it would be extremely important to measure
these cross sections in this energy range.

This talk is based on material obtained in collaboration with E.
Bratkovskaya, W. Cassing, M. Effenberger, A. Sibirtsev, S. Teis and G. Wolf.
I am very grateful to them for a fruitful collaboration and many critical
discussions.

\end{document}